# SPARSE SUPPORT RECOVERY WITH PHASE-ONLY MEASUREMENTS

*Yipeng Liu[1,2], Qun Wan[1], Fei Wen[1], Jia Xu[2], Yingning Peng[2]*

[1]Department of Electronic Engineering, University of Electronic Science and Technology of China (UESTC), Chengdu 611731, China

[2]Department of Electronic Engineering, Tsinghua University (THU), Beijing, 10084, China

## ABSTRACT

Sparse support recovery (SSR) is an important part of the compressive sensing (CS). Most of the current SSR methods are with the full information measurements. But in practice the amplitude part of the measurements may be seriously destroyed. The corrupted measurements mismatch the current SSR algorithms, which leads to serious performance degeneration. This paper considers the problem of SSR with only phase information. In the proposed method, the minimization of the L1 norm of the estimated sparse signal enforces sparse distribution, while a nonzero constraint of the uncorrupted random measurements' amplitudes with respect to the reconstructed sparse signal is introduced. Because it only requires the phase components of the measurements in the constraint, it can avoid the performance deterioration by corrupted amplitude components. Simulations demonstrate that the proposed phase-only SSR is superior in the support reconstruction accuracy when the amplitude components of the measurements are contaminated.

# 1. INTRODUCTION

Compressive sensing (CS) is a new signal processing technique. It can reconstruct the signal with fewer randomized samples than Nyquist sampling with high probability on condition that the signal has a sparse representation (Candes and Wakin, 2008; Candes et al., 2006; Dohoho, 2006). In CS, support recovery refers to the problem of correctly estimating the position of the non-zero entries from a set of noisy measurements (Tang and Nehorai, 2010). Orthogonal matched pursuit (OMP) (Needell and Vershynin, 2009), basis pursuit (BP) (Chen and Donoho, 1999) and Dantzig Selector (DS) (Candes and Tao, 2007) are the major ways to recovering the sparse signal support.

Most of the traditional sparse signal support recovery methods are with the full information measurements (Needell and Vershynin, 2009; Chen and Donoho, 1999; Candes and Tao, 2007). But in practical it may not be able to make full use of all the information of the signal (Goldsmith, 2005; Prymek, 1991; Pozidis and Petropulu, 1997). The acquired measurements for signal support recovery may not contain all the information as it transmits, and the digital signal processor should do with the measurements with part of the information lost. amplitude information is such kind of information. In many electronic systems, the signal's amplitude information can be seriously contaminated by channel fading which occur approximately in long-term exposure to atmospheric turbulence (Goldsmith, 2005), automatic gain control (AGC) errors (Prymek, 1991), and images which are blurred by defocused lenses with circular aperture stops (Pozidis and Petropulu, 1997), etc.

Instead of trying to make up the amplitude components of the signal, here we propose to estimate the sparse signal support from the incomplete measurements with only the phase information which we name as phase-only measurements. Different from the classical sparse signal reconstruction methods in

CS, the proposed sparse signal support reconstruction from phase-only measurements can be formulated as convex programming. It minimizes the L1 norm of the estimated signal to encourage the sparse distribution, while the uncorrupted random measurements' amplitudes with respect to the reconstructed sparse signal are restricted to nonzero. Monte Carlo simulations show that the proposed method has obvious support recovery accuracy enhancement.

In the rest of the paper, the sparse signal model is provided in section II. Section III gives the proposed sparse support recovery with phase-only measurements; In Section IV, the performance enhancement of the proposed method was demonstrated by numerical experiments; Finally Section V draws the conclusion.

## 2. SPARSE SIGNAL MODEL

Considering a complex signal $\mathbf{r} \in \mathbb{C}^{N \times 1}$ can be expanded in an orthogonal complete dictionary with the representation as

$$\mathbf{r}_{N \times 1} = \mathbf{\Psi}_{N \times N} \mathbf{x}_{N \times 1}, \qquad (1)$$

when most elements of the vector x are zeros, the signal r is said to be sparse. When the number of nonzero elements of $\mathbf{x}$ is $K$ ($K \ll N$, $K > 0$), the signal is said to be $K$-sparse. The support of the sparse signal $\mathbf{r}$ is the position of the nonzero elements in the sparse vector $\mathbf{x}$. we use the vector $\mathbf{s}$ to denote the support with the element:

$$s_i = \begin{cases} 1, & \text{for } x_i \neq 0 \\ 0, & \text{for } x_i = 0 \end{cases}, \quad \text{for } i = 1, 2, \cdots, N, \qquad (2)$$

Here we denote the operation to get the support $\mathbf{s}$ from the sparse signal $\mathbf{x}$ as:

$$\hat{\mathbf{s}} = \text{supp}(\hat{\mathbf{x}}, K), \qquad (3)$$

In CS, instead of measuring the signal directly by Nyquist sampling, a random measurement matrix $\mathbf{\Phi} \in \mathbb{C}^{M \times N}$ is used to sample the signal. In matrix notation, the obtained random sample vector can be represented as:

$$\mathbf{y}_{M \times 1} = \mathbf{\Phi}_{M \times N} \mathbf{f}_{N \times 1}, \quad (4)$$

The measurement matrix $\mathbf{\Phi}$ should satisfy the restricted isometry property (RIP) which is a condition on matrix $\mathbf{\Phi}$ which provides a guarantee on the performance of $\mathbf{\Phi}$ in CS. It can be stated as (Candes and Wakin, 2008; Candes et al., 2006; Dohoho, 2006):

$$(1-\delta_s)\|\mathbf{y}\|_2^2 \leq \|\mathbf{\Phi}\mathbf{y}\|_2^2 \leq (1+\delta_s)\|\mathbf{y}\|_2^2, \quad (5)$$

for all *K*-sparse $\mathbf{y}$. The restricted isometry constant $\delta_s \in (0,1)$ is defined as the smallest constant for which this property holds for all K-sparse vectors $\mathbf{y}$. There are three kinds of frequently used measurement matrices. The hardware to random sampling can refer to (Laska et al., 2006; Laska et al., 2007). When the RIP holds, several classical algorithms can reconstruct the sparse signal (Candes and Wakin, 2008; Needell and Vershynin, 2009; Chen and Donoho, 1999; Candes and Tao, 2007).

Here we reformulate the random measurement vector $\mathbf{y} \in \mathbb{C}^{M \times 1}$ as:

$$\mathbf{y}_{M \times 1} = \mathbf{\Phi}_{M \times N} \mathbf{r}_{N \times 1}, \quad (6)$$

It can be divided into two separate parts as:

$$\mathbf{y} = \mathbf{y}_a \odot \mathbf{y}_p, \quad (7)$$

Where $\odot$ is the Hadamard product, and

$$\mathbf{y}_a = \mathrm{abs}(\mathbf{y}), \quad (8)$$

$$\mathbf{y}_p = \mathrm{phase}(\mathbf{y}), \quad (9)$$

are the amplitude and phase components of the random measurement vector respectively.

We assume the amplitude components of the sparse signal are corrupted by a random Gaussian perturbation as:

$$\mathbf{z}_{M\times 1} = diag(\mathbf{v}_{M\times 1})\mathbf{y}_{M\times 1}, \qquad (10)$$

where diag(**v**) is a diagonal matrix with its diagonal elements coming from the vector **v**. We restrict the elements of **v** being nonzero to keep that only amplitude information of the random measurement can be corrupted.

The corrupted measurement vector can also be divided into two separate parts as:

$$\mathbf{z} = \mathbf{z}_a \odot \mathbf{z}_p, \qquad (11)$$

where

$$\mathbf{z}_a = \text{abs}(\mathbf{z}), \qquad (12)$$

$$\mathbf{z}_p = \text{phase}(\mathbf{z}), \qquad (13)$$

are the amplitude and phase components of the corrupted measurements respectively.

As only the amplitude components are corrupted by the random perturbation **v**, the phase component of the random measurement, denoted as $\mathbf{y}_p$, is kept in the phase-only measurements, i.e.

$$\mathbf{y}_p = \mathbf{z}_p, \qquad (14)$$

In practice when the amplitude components of the sparse signal are damaged, we have to use the phase-only measurement $\mathbf{z}_p$ to recover the sparse signal.

## 3. THE PROPOSED PHASE-ONLY SPARSE SIGNAL RECOVERY

There are several traditional methods to obtaining the support of the sparse signal. BP has almost the same performance as DS (James et al. 2009); and both BP and DS have better reconstruction accuracy than greedy algorithms, such as OMP, iterative thresholding (Blumensath and Davies, 2009), and so on. For a higher accuracy, BP is usually chosen to reconstruct the sparse signal. Here the random measurement matrix $\mathbf{\Phi}$ and the basis matrix $\mathbf{\Psi}$ are known is advance. We define $\mathbf{A} = \mathbf{\Phi}\mathbf{\Psi}$, and the BP can be formulated as:

$$\min_{\mathbf{x}} \|\mathbf{x}\|_1 \\ \text{s.t. } \mathbf{z} = \mathbf{A}\mathbf{x} \quad (15)$$

(15) is a second-order cone programming (SOCP). It can be solved by many software (Sturm, 1999).

When the amplitude information is seriously destroyed, we can discard the amplitude part of the measurements. As the phase-only measurements are used to recovery the sparse signal, the BP with phase-only measurements can be modeled as:

$$\min_{\mathbf{x}} \|\mathbf{x}\|_1 \\ \text{s.t. } \mathbf{z}_p = \mathbf{A}\mathbf{x} \quad (16)$$

(16) is named as phase-only basis pursuit (POBP), for it uses only the phase components of the measurements in the BP algorithm. Similarly, (15) is named as standard basis pursuit (SBP).

The BP was designed for the full information measurements. When nearly only phase information is able to use, the measurements would mismatch with the algorithm, and the recovery performance would be deteriorated. To find a more suitable algorithm from the phase-only measurements, we discharge the constraint $\mathbf{z}_p = \mathbf{A}\mathbf{x}$ in (16) where the mismatch is originated.

Here the amplitudes of the random measurements $\mathbf{y}$ can be stated as:

$$\mathbf{y}_a = \text{diag}(\mathbf{y}_p^*)\mathbf{y}. \quad (17)$$

Considering that the phase components of random measurements are the same with the corrupted measurements', we can get:

$$\mathbf{y}_a = \text{diag}(\mathbf{z}_p^*)\mathbf{y}. \quad (18)$$

Assuming the estimated sparse vector is **x**, we can get:

$$\mathbf{y}_a = \text{diag}(\mathbf{z}_p^*)\mathbf{A}\mathbf{x}. \qquad (19)$$

The amplitude vector $\mathbf{y}_a$ is non-negative. Here we let μ denote the minimum element of the amplitude vector $\mathbf{y}_a$. Thus we have:

$$\text{diag}(\mathbf{z}_p^*)\mathbf{A}\mathbf{x} \succeq \mu\mathbf{1}. \qquad (20)$$

where **1** is the $M \times 1$ vector with all its elements being ones. Considering the probability that the minimum element μ is zero is zero (Grimmett and Stirzaker, 2001):

$$\text{Prob}(\mu = 0) = 0. \qquad (21)$$

It is no bother to reformulate (20) as:

$$\text{diag}(\mathbf{z}_p^*)\mathbf{A}\left(\frac{1}{\mu}\mathbf{x}\right) \succeq \mathbf{1}. \qquad (22)$$

In support recovery of the sparse signal, the relative amplitude of the elements of the vector **x** can give out the support vector **s**, i.e. A constant multiplication by the vector **x** would have no influence on support recovery. Thus we can assume μ = 1, and reformulate (22) as:

$$\text{diag}(\mathbf{z}_p^*)\mathbf{A}\mathbf{x} \succeq \mathbf{1}. \qquad (23)$$

Combining (23) with the sparse constraint, we can get the proposed sparse signal reconstruction model as:

$$\begin{aligned} \min \ &\|\mathbf{x}\|_1 \\ \text{s.t.} \ &\text{diag}(\mathbf{z}_p^*)(\mathbf{A}\mathbf{x}) \succeq \mathbf{1} \end{aligned}. \qquad (24)$$

As only phase components of the measurements are utilized in the constraint, (24) is named as phase-only sparse signal recovery (POSSR). The phase-only measurements just match the constraint (23), for it also requires and only requires the phase-only measurements. (24) is also an SOCP, and can be solved by SeDuMi (Sturm, 1999). After getting the **x** from (24), the support vector *s* can be obtained by (3).

## 4. SIMULATION

In this section we present simulation results that demonstrate the performance of the proposed algorithm. Without loss of generality, we set **Ψ = I,** i.e. the signal is sparse with respect to the canonical basis of the Euclidean space. The length of the sparse signal is $N = 100$. The mean and variance of the random vector **v** are set to 1 and 0.5 respectively. The measurement matrix has i.i.d. coefficients which are drawn from a standard normal distribution. In every simulation we regard that the support recovery algorithm is successful if and only if the estimated support is the same as the true support of the generated sparse signal. We define the success probability of support recovery (SPSR) as

$$\eta = \frac{\mathrm{T}(\hat{\mathbf{s}} = \mathbf{s}_{ture})}{L}, \qquad (25)$$

where **ŝ** is the support of the estimated sparse signal and $\mathbf{s}_{trure}$ is the support vector of the generated sparse signal; $\mathrm{T}(\Xi)$ means the number of times that the $\Xi$ happens in the total *L* times Monte Carlo simulations; and *L* is the times of Monte Carlo simulations in total. To make contrast, the SBP and the POBP and the proposed POSSR are used to recover the support of the sparse signal.

Figure 1 gives out the simulated SPSR of the three methods in contrast to the real sparse signal with the number of nonzero elements $K = 5$ and the length of the random measurement vector $M = 100$. It is obviously that the proposed POSSR achieves the best performance. The estimated nonzero entries of the sparse signal by POSSR can be easily distinguished and their positions are correctly corresponding to the ones of the real sparse signal; while the noise levels of the other two estimated sparse signals are too high to make out the nonzero entries.

Figure 2 gives out the simulated SPSR of the three methods when the sparsity $K$ increases from 1 to 10 with the length of the random measurement vector $M = 100$, and the number of Monte Carlo simulation is $L = 1000$. It shows that the reconstruction accuracy of POSSR is obviously higher than the other two methods.

Figure 3 gives out the simulated SPSR of the three methods when the number of the random measurements $M$ increases from 10 to 120 with the number of nonzero entries $K = 4$, and number of Monte Carlo simulation $L = 1000$. It shows that the SRSR of the SBP is higher than the other two from $M = 10$ to $M = 80$, and it reaches the top at $M = 80$. The highest rate is 76.2 %, which is a little lower for a robust sparse support recovery. However, the SPSR of the POSSR is monotonically increasing with the increase of number of measurements. Here it achieves its maximum SPSR = 86.1 % at $M = 140$, which is superior to the maximum SPSRs of the SBP and the POBP. And we can predict that the SPSR would be even higher when the number of the measurements is larger than $M = 140$. Besides, at $M = 80$ the SPSR of POSSR is 73.6 %, which is close to the corresponding SBP's SPSR.

## 5. CONCLUSION

This paper introduces a novel sparse support recovery method when part of the amplitude information of the measurements is lost. In the proposed method, sparse constraint in the form of minimization of *L1* norm of estimated signal is combined with the nonzero constraint of the random measurements' amplitudes with respect to the phase-only measurements. Simulations demonstrate that the proposed method achieves higher success probability of sparse signal recovery.

In the furture, the theoritical sufficient and nessessary condtions can be deduced to further solidate the performance of the phase-only measurement based sparse support recovery. Besides, the proposed POSSR is a convex programming. As we know that convex way has the higher accuurancy while greedy algorithm has the lower computational complexity, it is worthy of finding the corresponding greedy algorithm. Finally, the proposed POSSR estimates a single vector from a single snap, and it can be generalized to the Multiple Measurement Vectors (MMV) situation.

## 6. ACKNOWLEDGEMENT

This work was supported in part by the National Natural Science Foundation of China under grant 60772146, the National High Technology Research and Development Program of China (863 Program) under grant 2008AA12Z306, the Key Project of Chinese Ministry of Education under grant 109139, China National Science Foundation under Grant 60971087, China Ministry Research Foundation under Grant 9140A07011810JW0111 and 9140C130510D246, Aerospace Innovation Foundation under Grant CASC200904.

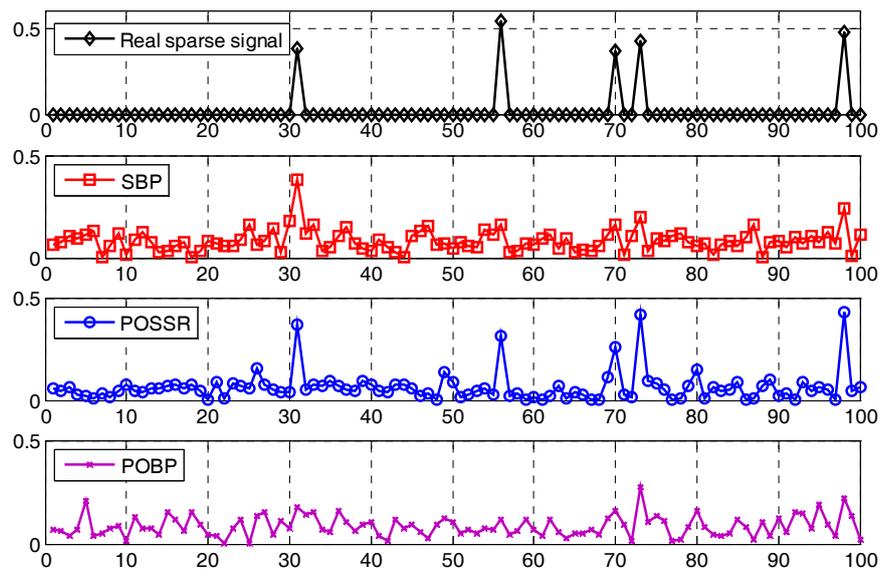

Figure 1. The sparse signal recovery of three methods in contrast to the real sparse signal with the number of nonzero elements $K = 5$, the length of the random measurement vector $M = 100$.

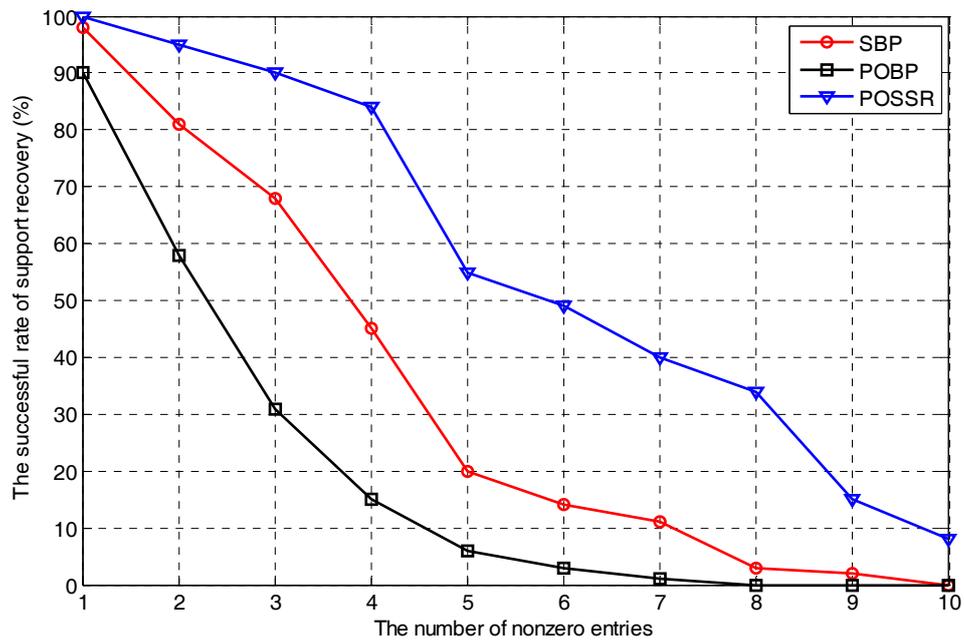

Figure 2. Simulated success probability of the three methods when the number of nonzero entries M increases from 1 to 10 with the number of the random measurements M = 100, and the number of Monte Carlo simulations L = 1000.

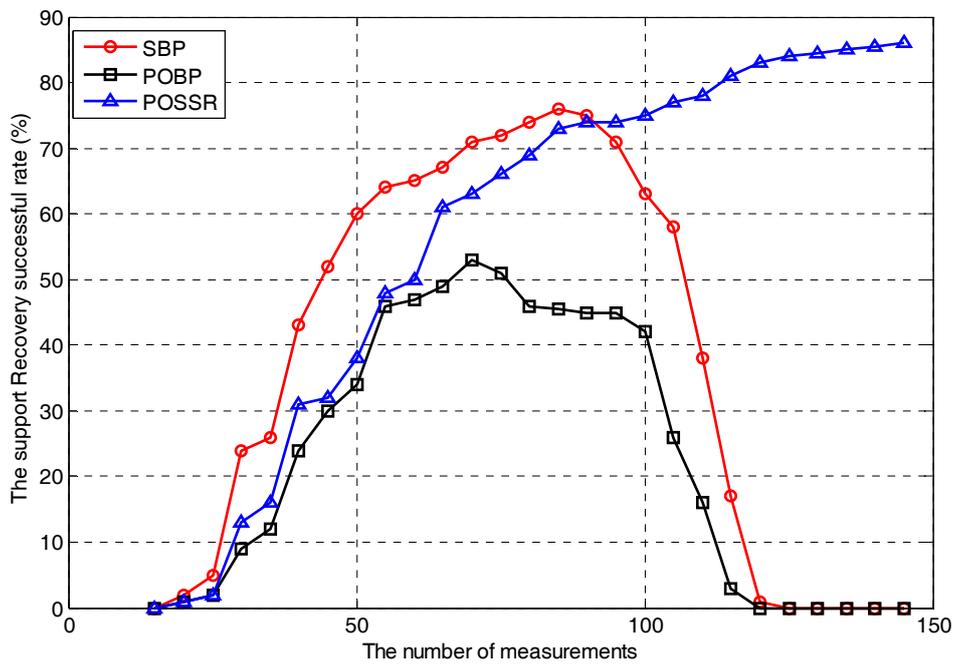

Figure 3. Simulated success probability of the three methods the three methods when the number of the random measurements M increases from 10 to 140 with the number of nonzero entries $K = 4$, and the number of Monte Carlo simulations $L = 1000$.